# Recent Innovations in Footwear and the Role of Smart Footwear in Healthcare- a Survey

Pradyumna G. R.[a], Roopa B. Hegde[a], Bommegowda K. B.[a], Anil Kumar Bhat[a], Ganesh R. Naik[b] and Amit N. Pujari[c, d]*,

[a] NITTE (Deemed to be University), Dept. of Electronics and Communication Engineering, NMAM Institute of Technology, Nitte-574110, Udupi, Karnataka, India.
[b] College of Medicine and Public Health, Flinders University, Adelaide, Australia
[c] School of Physics, Engineering and Computer Science, University of Hertfordshire, England, UK.
[d] School of Engineering, University of Aberdeen, Scotland, UK
pradyumna@nitte.edu.in, roopabhegde@nitte.edu.in, bgowda_kbl@nitte.edu.in, anilkumarbhat@nitte.edu.in, ganesh.naik@flinders.edu.au, amit.pujari@ieee.org

*Amit N. Pujari
**Email:** amit.pujari@ieee.org

**Author Contributions:** Study conception, data collection and draft manuscript preparation: Pradyumna G. R., Roopa B. Hegde, Bommegowda K. B., Anil Kumar Bhat, Ganesh R. Naik and Amit N. Pujari. All authors reviewed and approved the final version of the manuscript.

**Competing Interest Statement:** The authors declare no competing interests.

**Classification:** Healthcare.

## Abstract

Smart shoes have ushered in a new era of personalized health monitoring and assistive technology. The shoe leverages technologies such as Bluetooth for data collection and wireless transmission and incorporates features such as GPS tracking, obstacle detection, and fitness tracking. This article provides an overview of the current state of smart shoe technology, highlighting the integration of advanced sensors for health monitoring, energy harvesting, assistive features for the visually impaired, and deep learning for data analysis. The study discusses the potential of smart footwear in medical applications, particularly for patients with diabetes, and the ongoing research in this field. Current footwear challenges are also discussed, including complex construction, poor fit, comfort, and high cost.

## Significance Statement

This study presents the recent innovations in footwear, particularly the integration of smart technology, that have transcended traditional expectations. The focus on healthcare applications not only enhances user experience but also opens new avenues for preventive and personalised healthcare. The synergy between technology and footwear has the potential to revolutionize healthcare practices, offering proactive solutions for individuals across various demographics.

## Introduction

In recent years, technological advances in the Internet of Things (IoT) and wearable devices have penetrated the shoe industry, driving smart shoe design. Integration of electronic and mechanical components, controllers, and microprocessors has enabled the development of a shoe into a smart shoe. Using smart footwear (SF), personalised cell phone applications are also available for self-monitoring (1). SF can capture, monitor and record activities of daily living. Particularly, the use of intelligent sensor(s) enables footwear to act as a SF. As sensors can keep track of activities of daily living and can provide information about a range of vital characteristics such as: (foot) pressure points, posture correctness, body fatigue level, temperature in the footwear, number of steps completed, weight of a person, and real-time location. This information can be further automatically analysed and used for diagnosis. Hence, SF can provide personalised feedback to users, from measuring athletic performance to tracking fitness and evaluating health metrics. The efforts in this area seemed to be aimed at integrating technology to improve comfort, convenience, and good health (2).

The survey findings highlight how continuous research and technological progress are empowering the creation and advancement of SF for diverse applications. Driven by this capability, businesses are actively innovating in the SF sector to address the rising demand from consumers. Considering the recent surge in research and innovation specifically within the SF domain, this survey aims to pinpoint existing challenges in the design and development of SF. Additionally, it outlines future perspectives concerning SF technology. Further, the survey also provides fundamental technological elements of SF and explores how current technologies can be harnessed to streamline application processes. Thus, the article offers a concise overview of the roadmap for integrating modern technologies into footwear for improved lifestyles.

**Search strategy:** To search the relevant literature, a Scopus database with keywords "smart footwear," "IoT footwear," and "intelligent footwear" was used. In view of providing recent trends and developments in footwear technology, articles published between 2017 and 2023 applicable to healthcare are reviewed. The number of published articles included in the review are shown in Figure 1. The number of recent articles published on SF reflects increased research and innovation activity in this domain, perhaps motivated by increased consumer/business need for innovative SF.

**Organisation of the Survey:** Considering the importance of sensors in SF, firstly, sensors suitable for and commonly used in SF are explained with some detail. Then the survey begins with providing an overview of a) SF implementation/use in the real world and b) SF's main healthcare applications. Subsequently, the three main healthcare applications of SF 1) Performance tracking, 2) Patient Monitoring and 3) Detection and Recognition are surveyed in-depth. After surveying SF's main healthcare applications, a brief overview of 'the various types of smart footwear available on the market' is provided. Finally, key observations are drawn from the reviewed literature followed by future perspectives and conclusions derived.

SF can be broadly categorized into a) passive, which derives its energy from mechanical (limb) movements, and b) active, which works based on sensors. Notably, the development of SF often involves a fusion of both these categories, resulting in hybrid models that utilize elements from both passive and active technologies to enhance their overall functionality and performance. Such smart footwear can be used for various applications. The design and development of smart footwear involve multiple aspects, as depicted in Figure 2. The listed applications are related to human health monitoring. However, SF's applications are not only limited to human health monitoring but also used as a tracking and support device.

Sensors are electronic devices that in SF context, typically record physical, analog parameters (such as pressure, temperature, movement) and convert them into digital form which can be processed to draw insights into the activities of daily living (e.g. number of steps, feet pressure points/gait etc.). A wide variety of sensors are available, and the choice of a sensor depends on the SF requirements. Table 1 lists sensors which are suitable for SF and hence are commonly used in the SF design.

**Overview of smart footwear and its applications in healthcare**

In recent years, the utility of SFs has been commonly found in performance tracking, health monitoring, and the detection of specific disorders, as depicted in Figure 3. This section provides an overview of the state-of-the-art methods used in these applications.

The design of SF involves integration of sensor into footwear followed by data acquisition, and data analysis, processes. Figure 4 illustrates these processes. SF also includes an inbuilt observing circuit that gives footstep count, weight count, walking speed, travel mapping, distance count, and pro-health tips (14). Once the sensor collects the data, this data needs to be sent to the processing system for data analysis. This requires controller boards typically microcontrollers such as Arduino, Raspberry Pi, etc. Nowadays, wireless communication is commonly used over wired communication to carry out the data transfer. Automated data analysis and results can be displayed in smart footwear, connected to an App (14). Finally, a decision is made based on the analysis report which can be generated automatically. Analysis reports can enable remote assistance. Considering the implications for a person's health and SF's input in aiding medical decision making, data analysis employed by the SF needs to be clinically validated, robust, safe and respect patient (personal) data privacy.

*Application 1: Performance tracking*

This section focuses on the state-of-the-art methods reported for tracking the visually impaired, measuring athlete performance, and monitoring older adults and soldiers.

Globally, at least 2.2 billion people have a vision impairment or blindness (15). Visually impaired find it difficult to navigate from one place to another place independently. SF can assist the visually impaired in navigating safely. Use of ultrasonic sensors with Arduino UNO (16, 17,18, 19), Node-MCU (20), Raspberry Pi (21), and microcontroller (22, 23) can be found in the SF literature. These footwear can detect obstacles and guide the user to avoid them. Integrating GPS tracking with wireless charging systems, as proposed by Thanuja et al. (21), can provide real-time location information about the user, allowing their friend or family members to be informed when needed. An extensive study conducted by Hersh et al., (24) discuss various types of wearable devices designed for the travelling aid of visually impaired. They (24) describe SF with ultrasonic obstacle detection and additional functions such as a water detection sensor to detect wet floors, a 3-axis Accelerometer, and a 3-axis gyroscope for falls. However, this review focused mainly on design issues. In another study, secure movement with low faulty error was observed when the visually impaired were tested with smart footwear with obstacle detection, wet sensing, and fall detection functions (25). In addition, smart footwear was designed to facilitate blind people in finding their way back to their destination in case they miss their route (26).

In addition to tracking, physical parameters such as speed, pressure, and acceleration can be measured with sensors integrated into footwear. Also, physiological parameters such as oxygen saturation level, body temperature, and blood pressure can be measured. This would help to understand the training level and performance of the athlete. An investigation by Knopp et al. (27) revealed that the running economy could be improved by using smart footwear runners. A recent study by Gulas and Imre (28) mentioned that along with improving the technological aspects, it is necessary to look into the textile quality and material used for developing smart footwear. Another study reported the importance of smart socks and in-shoe systems for sports and medical applications (29). Sports injury is an unavoidable circumstance, and this requires rehabilitation for a speedy recovery. Also, gait analysis may be suggestable to detect any severe injury. A study by Lianzhen and Hua (30) evaluated athlete performance rehabilitation by providing Internet of Health Things (IoHT) guidance.

Further, porous polymer dielectric was used for capacitive sensors to improve the pressure sensing of capacitive sensors for measuring heel pressure while running and walking of athletes (31), yielding considerably greater signal changes than the other dielectric materials. Barratt et al. (32) tested the reliability of plantar pressure measurement. They validated it by comparing the performance of the smart insole with PedarX and concluded that a wireless insole might be more practical than a wired system. An article by Beckham (33) described features of next-generation athlete shoes that allow users to customize the look, fit, and responsiveness of kicks. Also, such shoes can transmit data to the cloud and will receive feedback to fine-tune the workout by connecting to Facebook or Twitter. Though there is a need for automated continuous monitoring and assistance for athletes, another major requirement lies in monitoring older adults.

Worldwide elderly population is growing and according to WHO, it is likely to reach 2.1 billion by 2050 (34). With ageing comes the risks of age-related health problems. For example, WHO reports approximately 684,000 fall-related deaths every year worldwide, which are likely to increase with the ageing population (35). Smart footwear can help in minimising falls by monitoring older adults. An article by Schiltz (36) indicated that smart footwear with GPS devices benefits older adults, especially those with Alzheimer's or cognitive disorders. LI (37) identified the core risk factors to be addressed while designing footwear for fall detection and proposed an optimised design. A possible smart footwear design for location tracking and health monitoring of older adults with sensors, GPS, and RFID technology was demonstrated by Cheng et al. (38).

SF can be particularly useful where the user does not have access to or support of a carer. In this direction, smart footwear designs and proposals can be found in the literature (39, 40, 41, 42). The state-of-the-art methods addressed tracking of elderly and also monitoring walking speed (39). Older adults can be safeguarded either by carers or by using technology-driven systems.

Defence systems, including the army, navy, and air force, face the issue of tracking the soldiers. Smart footwear is a feasible solution for tracking soldiers. Smart footwear-based remote monitoring and location tracking of soldiers can be found in the literature using LoRa (43) for longer-distance communication, LoRa, and mobile applications (44). The reported articles addressed location tracking and lack of monitoring of physical or physiological parameters. An extensive study in 2018 by Friedl (45) focused on all possible wearable devices, including smart footwear, for monitoring soldiers remotely. This study brought out the challenges and future directions. Both health monitoring and position tracking of soldiers are addressed using various sensors, PIC microcontrollers, and GPS (46). An additional panic button was provided at the soldier's end to ask for help from the base camp.

The above section provided an overview of studies addressing performance tracking used in four applications: a) vision impairment, b) athletics, c) elderly and d) defence. Figure 5 depicts the proportion of articles published pertaining to each application category.

### Application 2: Patient Monitoring

In recent years developments in wearable technology have enabled remote monitoring of patients in healthcare centres and houses. Monitoring may include daily activity recognition, monitoring patients with walking issues or monitoring patients with specific conditions such as diabetes, cardiac issues, gait disorder, etc.

Human activity recognition (HAR) is extensively used in various applications such as day-to-day activities, health monitoring, fitness tracking, and monitoring of the physically impaired. Wearable devices available for HAR smart footwear are gaining popularity in recent years. An investigation by Truong et al. (47) revealed that wrists or feet are suitable for remote HAR. Hence, footwear is a feasible solution for recording human activity and analysing it remotely. For HAR application, pressure, inertial, or both sensors can be used. However, it was found that inertial sensors are reliable for the recognition of dynamic activities, while pressure sensors are reliable for stationary activities (48). Hence, using both inertial and pressure sensors is suggested for efficient outcomes. Recognition is a task that requires decision-making by processing sensor data. Machine learning (ML) and deep learning (DL) are identified as most suitable for recognition tasks. The use of DL for HAR can be found in the literature (49, 50). Plantar pressure measurement is an important aspect of clinical studies; hence, it is essential to calibrate the plantar pressure value given by the sensor before integrating it with the footwear. An experimental evaluation by Kakarla (51) disclosed that ML can model smart footwear with required pressure measurement.

Further, a random forest (RF) algorithm was employed by Ren (52) for daily activity recognition by measuring plantar pressure. Capacitive sensors showed promising results for measuring plantar pressure (53, 54), allowing HAR. Another study by Pham et al. trained a DL algorithm using accelerometer data obtained from smart footwear and reported an accuracy of 93 % (55). Image-based 3D analysis was conducted to test the effectiveness of such smart footwear on healthy volunteers (56). Modern smartphones already come with an array of inbuilt sensors, and a study by Dogan et al. (57), demonstrated the use of smartphone sensors for HAR by employing DL. A daily activity recognition system can be used for monitoring healthy people. Disease-specific monitoring however requires several parameters specific to the disorder. Hence, the SF specific to particular conditions may involve additional hardware and software tools. One such widespread condition that requires regular monitoring is diabetes.

Diabetes is a chronic disease that damages the heart, blood vessels, eyes, kidneys, and nerves. About 422 million people worldwide have diabetes, the majority living in low-and middle-income countries, and 1.5 million deaths are directly attributed to diabetes each year (58). Within a diabetic population, diabetic foot ulcer (DFU) is a life-threatening complication. A recent study reported that the mortality rate due to DFU is high and is about 50 % in five years (59). Hence therapeutic footwear is commonly recommended for early detection and diagnosis. However, this requires continuous monitoring of such patients. Various precursors and risk factors of DFUs include joint contractures, Arthritis, and callus formation. A pronged approach, including patient education, appropriate footwear selection, telehealth, and proactive surgical interventions, are essential measures for preventing new and recurrent DFUs (60) as depicted in Figure 6. Smart footwear is a feasible solution to monitor patients remotely, helping manage DFU more effectively.

DFUs are a common complication of diabetes, often resulting from nerve damage or poor blood circulation. Hence, managing DFUs is crucial due to their potential to lead to severe infections. Smart footwear, integrated with advanced technologies, has emerged as a promising tool in early detection, monitoring and managing diabetic foot complications. Hence, several innovative approaches are emerging to monitor DFU by integrating sensors in footwear. Moulaei et al. (61) and Altaf et al. (66) showed the accurate measurement of pressure, humidity, and temperature of patients' feet and sending this data to their smartphones by the Bluetooth module. A similar study by Sousa et al. (62) developed a need-based SF to monitor plantar pressure. However, mobile-based a plug-and-play device ("Dia Shoe") developed by Kularathne et al. [63] efficiently measured the temperature, humidity, weight and step count of the patient through a mobile application. A prototype SF model integrated with foot pressure and blood flow monitoring system having wireless data transmission (64) showed promising results for plantar pressure measurement (64). A proof-of-concept study (65) paired a smartwatch and SF consisting of eight pressure sensors for monitoring plantar pressure and alerting the individual through the watch. A flexible printed circuit board (PCB) based insole design (67) incorporated eight capacitive sensors sending data to a personal computer (PC) using a Bluetooth module enabling continuous monitoring of plantar pressure for real-time data analysis and evaluation. A study (68) recorded data associated with patients' preference toward footwear, insole design and quality-of-life-related information for further analysis and declared that patient-centric SF design is the key point for the therapeutic outcomes. A feasibility study (69) for measuring plantar pressure revealed that insole optimisation holds promise to evaluate the effectiveness of SF to monitor diabetic neuropathy. The temperature of the skin surface varies in the presence of an ulcer or wound on the surface. This fact was utilized (70, 72) to design SF integrating footwear and temperature sensors for early detection of foot ulcers in diabetic patients. Additionally, Zhang et al. studied deformation in the foot due to diabetes, specifically in women, by measuring plantar pressure (71).

Designing smart footwear involves several features, such as the selection of sensors, number of sensors, placement of sensors, communication mode, selection of controllers and processors, etc. Study (72), provides a detailed description on designing aspects of smart footwear. However, the choice of textile materials used in footwear can alter its efficiency (73, 74). Smart footwear must satisfy user needs and simultaneously also meet aesthetic/perception thresholds (75). Sometimes older adults may be hesitant to appreciate and accept smart/modern technology. Sometimes older adults may be hesitant or may not appreciate and accept smart/modern technology. This might cause mental disturbance during the usage of such systems. However, a study by Macdonald has shown that ML models can be used to monitor psychosocial factors to indicate a preference towards using smart footwear and adapting to the technology (76). Designing customised yet satisfactory and effective smart footwear for monitoring diabetic neuropathy is complex. A statistical framework approach for evaluating smart insoles has been used (77), and it was found that computer-aided designing of SF, using computer-aided design and computer-aided manufacturing (CAD-CAM), achieves better offloading performance than the traditional shape-only-based approach. Hence, developing innovative tools to support the design and manufacturing of customised footwear for people with diabetes is a crucial step and a CAD-based platform can help achieve this. Future research is needed to develop and optimize hardware tools and implement further design modules, namely insole-outsole, material selector, and valuator. However, the requirements of SF design may change according to the specific application or user group it is designed for, example, Parkinson's disease (PD), stroke, any injury, etc.

Ageing can contribute to increased foot pain after prolonged daily activities. This may lead to foot supination (body weight falling on the outer edge of the feet) or overpronation (weight falling inward). To avoid further complications caused by foot supination or overpronation, maintaining appropriate posture while walking, standing, or carrying out any daily activities is important. Unaddressed, these foot overcorrections can otherwise worsen and may develop foot disorders called gait disorders. Rehabilitation can be effective in treating posture disorders and biomechanics plays an important role in understanding posture (78). In recent years, smart footwear has enabled assessing gait and mobility disorders using biomechanics parameters (78). A study reported that a person's gait pattern is strongly influenced by age, personality, and mood (79). Foot impairments may affect daily activities and hence affect the quality of life. Several methods were proposed for developing smart footwear as an assistive device in case of gait disorder (80, 81, 82, 83, 84, 85). Further, the material used in footwear directly affects users' comfort, and a soft material-based smart insole could also provide equally good results (86). Wu et al. (87) demonstrated that analysis of gait disorders can be successfully achieved in real-time.

Gait analysis involves the evaluation of several parameters such as muscle strength in relation to limb activities, spatiotemporal joint kinematics, joint force, pressure distribution, plantar pressure, etc. Evidence exists for using an inertial sensor to detect events by measuring spatiotemporal parameters (88). A similar approach can be found using multi-cell piezoresistive sensors, inertial measurement, and logic units for measuring stride length, velocity, and foot clearance employing the ML technique (89). Various methods are reported for measuring CoP trajectories and Kinematic parameters (90), estimating progression angle while walking (91), monitoring alcohol-impaired gait (92), measuring kinematic and kinetic parameters (93), and determining plantar pressure (94). A smart insole PODOSmart system (95) measures spatiotemporal and kinematic gait parameters using wireless sensor technology and microprocessors.

Further, a pilot study was conducted to evaluate the performance of smart footwear with Tread Port virtual reality system for providing gait training (96). The study concluded that there is a requirement to boost the efficiency of smart footwear. Nowadays, custom-made capacitive sensors such as lightweight textile-based (97), Flexible Porous Graphene based (98), and Polydimethylsiloxane Composite material based (99) can be found in smart footwear for monitoring posture by measuring foot pressure. Another study (100) tested the efficiency of textured and prefabricated insoles by inserting them in medical and sports shoes, obtaining a significant change in reach distance compared to barefoot. A detailed technology evolution addressing gait disorder can be found in a review article (101), covering various types of smart wearables and described their benefits.

Several research groups have addressed different patient monitoring applications by designing SF for specific applications discussed in the present section. Technical insights and an overview of these approaches is provided in Table 2 for the reader's quick reference.

***Application 3: Detection and Recognition (Classification of Disorders)***
Human data, such as physical and physiological parameters, can be captured using sensors for further analysis to aid decisions during diagnosis or treatment. Sensor data recorded via SF, are typically time series, and extracting required information using manual timestamping and analysing can be time consuming, resource intensive and challenging. Hence, detection and analysis of sensor data recorded via SF, for identifying underlying patterns to categorise the type of disease accurately is mostly done automatically. For the automation of sensor data analysis and detection, Machine Learning (ML) or Deep Learning (DL) has been recently employed by the researchers. Different types of ML and DL algorithms used for healthcare applications are shown in Figure 7. Commonly used algorithms in reported articles are highlighted in different colours for clarity.

Jain et al. (102) classified accelerometer data for three walking activities using statistical methods. Another approach by Aqueveque et al. yielded promising gait analysis results by segmenting and analysing pressure data acquired from custom-made capacitive sensors made of two superimposed flexible copper films (103). Classification of gait patterns was addressed using the traditional supervised ML approach (104, 105, 106, 107, 108, 109) and the DL approach (110, 111, 112, 113114). Further, smart footwear data was used for training NN to recognise foot pronation and supination (115). A study by Moore et al. (116) compared the performances of several ML algorithms to predict strike angle and classify the strike pattern, classifying the strike pattern and getting misclassification only in the case of the mid-foot strike pattern. Fall is an unintentional event leading to injury, and it can happen to normal people or people with gait disorders or during rehabilitation. Hence, many smart footwear designs can detect falls (117, 118, 119), reporting promising results for practical scenarios.

Many researchers concentrated on designing and developing smart footwear for gait analysis. However, the analysis result depends on the type of sensors used, data acquisition bandwidth, sampling, and visualization. An experimental study can be found on the selection of bandwidth and sampling frequency for accurate classification of gait patterns (120). An investigation by Codina et al. (121) demonstrated that wireless technology and mobile application integration into smart footwear could be used for gait analysis and recovery speed monitoring after hip surgery and can also be used for fall detection. Another method developed by Sudharshan et al. (122) employed a smartphone-based application using a decision tree algorithm to classify walking patterns obtaining a classification accuracy of around 92 %. The design included five pressure sensors, four vibrators, and a Bluetooth transmission. Walking pattern analysis and hence assistance are required for patients suffering from Parkinsons' Disease (PD). A special footwear design employing closed-loop sensing to assist in the rehabilitation process of PD patients by analysing the walking pattern was presented by Cai et al. (123). In addition to the aforementioned smart footwear designs, a study showed the possibilities of smart socks to measure plantar pressure and analyse gait patterns (124). Suitability of smart footwear for recognition of many disorders and identification of body parameters such as heart rate estimation (125, 126), neuromuscular disease (127), postoperative outcome predictors (128), human behaviour classification using pneumatic actuators (129) and knee abduction moment prediction (130) were being experimented by the researchers.

The present section provided the details of SF data analysis for recognizing and classifying specific disorders. This was either achieved by using statistical tools or by employing ML algorithms. Table 3 provides an overview of the methods that reported disease or pattern classification. The table also highlights the approaches' outcomes, enabling the researchers to identify the gaps. A major issue that can be identified from the table is the smaller number of participants involved in the experiments limiting the confidence and extrapolation of the results.

***Various types of smart footwear available on the market***

Generally, younger generations are more inclined towards technology-incorporated products. Smart footwear is one such product that provides personalized feedback. Perhaps this is one of the reasons that smart footwear expanded at a robust compound annual growth rate (CAGR) of 22.7 % during the forecast period 2023-2033. The market is expected to hold a share of USD 269 million in 2023, while it is anticipated to surpass USD 2.1 billion by 2033 (136). The current section attempts to provide the details of the available smart footwear on the market. Figure 8 depicts the smart footwear companies on the global market. The Figure 8 offers popular manufacturers of smart footwear and their functions.

The key feature for market growth is increasingly higher expenditure on footwear by consumers. Companies are competing to meet the growing demands of consumers. However, design complexity and cost seem to be proportional to the features of the SF, as shown in Figure 9. It is evident from Figure 9 that computational complexity increases with an increase in the functions. This leads to hardware and software complexity resulting in increased cost. However, the trend towards SF is growing yearly; hence, manufacturers are continuously upgrading the technology per the growing demand. The details of the SFs available in the market are provided in Table 4.

### *Observations and future perspectives*

Smart footwear has been revolutionizing the future of footwear with the introduction of technology in product design and development. From monitoring physical health attributes to evaluating health benefits, smart footwears can enable wearers to receive personalized feedback. Customised smart footwears equipped with sensors, controllers, and processors assist and classify patterns. Based on the current survey, a few key observations are made.

Integration of sensors in smart footwear is continuously evolving, providing detailed health and fitness data for analysis. This includes tracking diverse health metrics, such as blood oxygen levels, heart rate, and energy expenditure. ML and DL algorithms enhance the accuracy and predictive capabilities of SFs, allowing for early detection of health issues based on changes in walking patterns or other metrics. Essentially, the ability of SF to become an efficient assistive device has been designed for individuals with disabilities, such as those who are visually impaired. Features like obstacle detection, GPS tracking, and machine learning-based object identification were used to help individuals navigate their environment more independently. Integration of wireless communication technologies also allows real-time alerts and updates to be sent to a user's smartphone or other devices. Expansion of SF application is also observed in the healthcare applications, particularly for patients with diabetes and walking disorders. It has been demonstrated that footwear monitoring pressure, temperature, and humidity could help prevent and manage conditions like Diabetic Foot Ulcer.

Further, optimising the sampling rate for data analysis has been reported to improve the accuracy. The development of adaptive sampling algorithms was reported. This allows smart footwear to adjust its data collection based on the user's activity level. Also, the use of smartphone Apps to collect and analyse data from SF has been proposed in many studies. This will allow users to easily track their health and fitness progress, receive alerts and recommendations, and share data with their healthcare providers. Improved and user-friendly designs for long-term usage were also reported. However, battery life was the major concern in many intelligent SF designs. This was also addressed by employing energy harvesting and storing techniques based on walking, to power the footwear's sensors and other electronic components. As shown in Figure 9, incorporating all these features increases the cost of the SF. However, the need is for cost-effectiveness, making it accessible to a wide range of consumers.

Though many SFs are available on the market, many studies can be found in the literature highlight the fact that design, and development of SF requires further research attention. Important avenues for future SF research can be summarised as follows. As sensor technology continues to evolve, opportunities to develop new types of sensors that can monitor a broader range of health and fitness metrics increase. This could include sensors that can detect changes in blood flow, muscle activity, or other physiological parameters. Another aspect is energy harvesting and storage. There is significant potential for research into more efficient energy harvesting and storage methods. This could involve developing new materials or designs that capture more energy from walking and/or creating more efficient energy storage systems to power the footwear's electronics for longer periods. More research is required on low-cost, energy-efficient, multi-parameter integrated sensor measurement units that can be used in SF. Clinically usable SF needs to be a decision support system for disease diagnosis. Hence, there is a need for ML and artificial intelligence (AI) methods to convert footwear to intelligent footwear. The use of these algorithms in SF is still relatively new, and there is much potential for further research in this area. This could involve developing algorithms that can more accurately predict health issues based on sensor data or creating AI systems that provide personalized advice.

Also, there is a need for more research into how smart footwear can be used to assist individuals with disabilities. This could involve developing new features or technologies to help these individuals navigate their environment or studying how smart footwear can be integrated into existing assistive devices or systems. Further, designing SF that can prevent or manage specific medical conditions or developing new types of therapeutic footwear that can deliver targeted treatments is in demand. As SFs become more common, research is needed into how to improve the user experience. This could involve studying how to make smart footwear more comfortable, user friendly and stylish and/or researching how to make the data collected by smart footwear more accessible and valuable to users. With the increasing amount of personal health data being collected by smart footwear, a question is how to protect this data. Hence, attention should go towards developing new encryption methods, privacy protocols and studying how to educate users about the importance of data security. As devices become increasingly smart and interconnected, researchers can explore how smart footwear can be integrated with other devices, such as smartphones, smartwatches, or home automation systems. Ultimately, sustainability is an essential consideration in product design, and there would be research and innovation opportunities in developing smart footwear using sustainable materials and/or manufacturing processes. Consumers or end-users care about the cost of any such devices. Hence, this opens a new avenue to manufacture cost-effective SFs by understanding manufacturing techniques and production processes while maintaining quality and performance.

**Conclusions**

In conclusion, smart footwear represents a significant advancement in wearable technology, offering promising applications in health monitoring, assistive technology, and medical treatment. Integrating advanced sensors, energy harvesting systems, and machine learning algorithms can revolutionise personal health management and improve the quality of life for individuals with disabilities. However, further research is needed to overcome current challenges, including improving the comfort and design of smart footwear, reducing costs, and ensuring data privacy and security. Future research should also explore the integration of smart footwear with other smart devices, using sustainable materials and manufacturing processes, and the development of more efficient energy harvesting and storage systems. As the field of smart footwear continues to evolve, it holds a promise of transforming our understanding of personal health and fitness and opening up new possibilities for assistive and therapeutic interventions.

**Acknowledgements**

We would like to express our sincere gratitude to NITTE (Deemed to be University), University of Hertfordshire, University of Aberdeen, and Flinders University for their valuable support during the writing process of the review article.

**Figures and Tables**

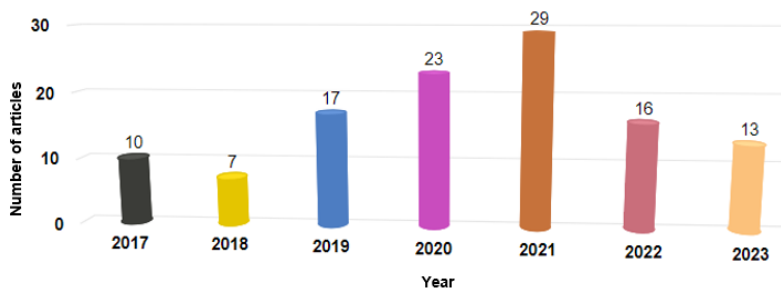

**Figure 1.** Articles included in the review.

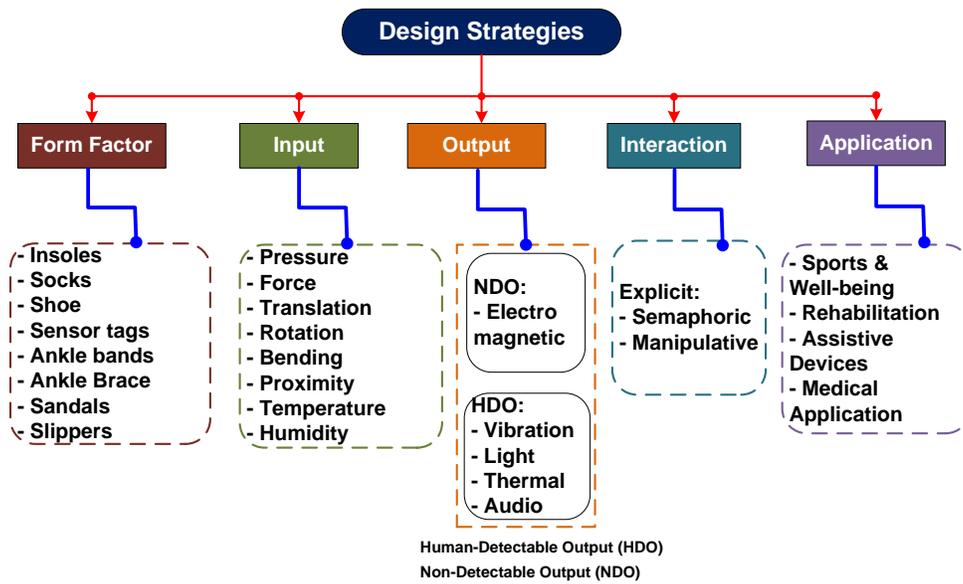

**Figure 2.** Various design aspects of smart footwear.

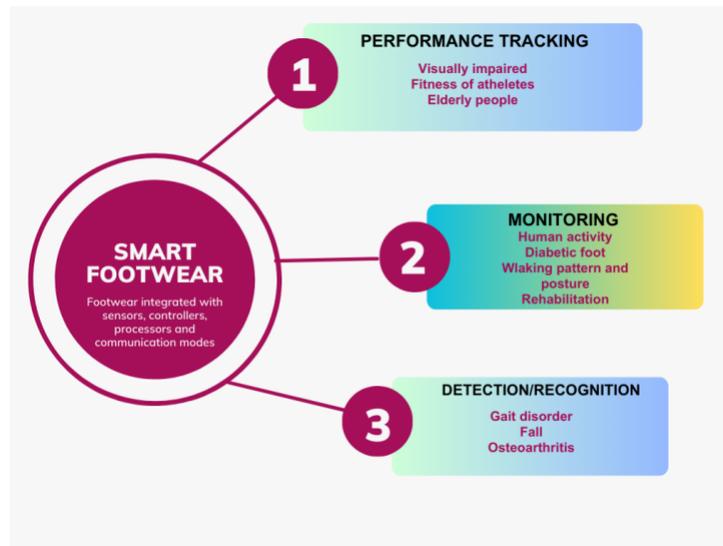

**Figure 3.** Main applications of smart footwear.

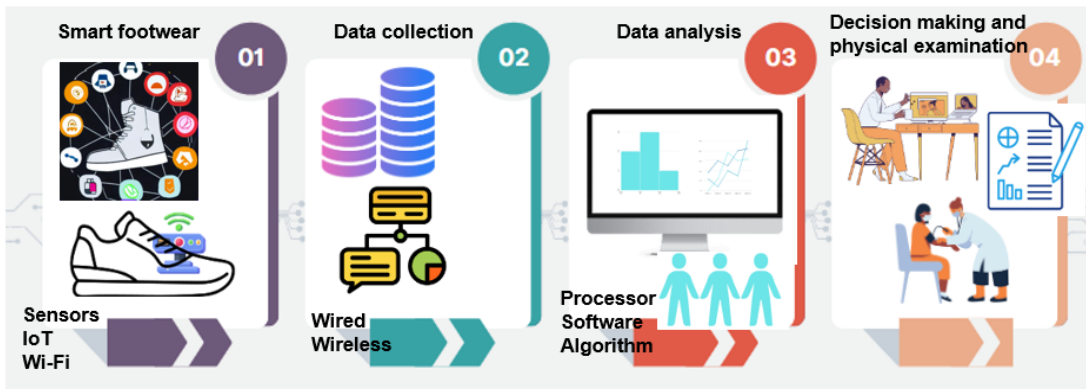

**Figure 4.** Generic workflow of IoT-based smart footwear.

● **Vision impaired** ● **Athletes** ● **Elderly** ● **Soldiers**

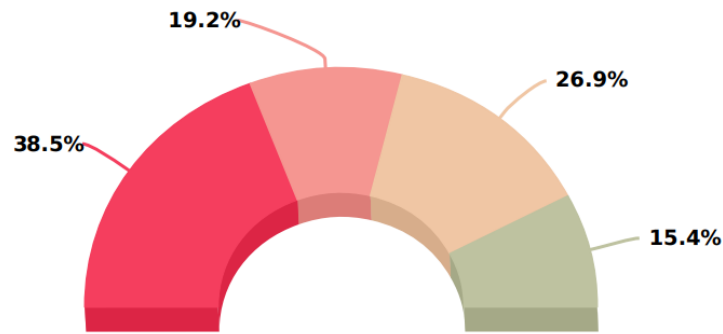

**Figure 5.** Number of articles addressing performance tracking.

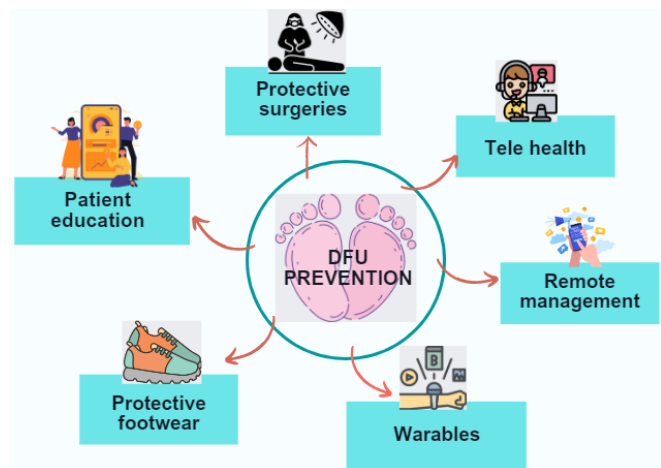

**Figure 6.** Preventive actions for DFU.

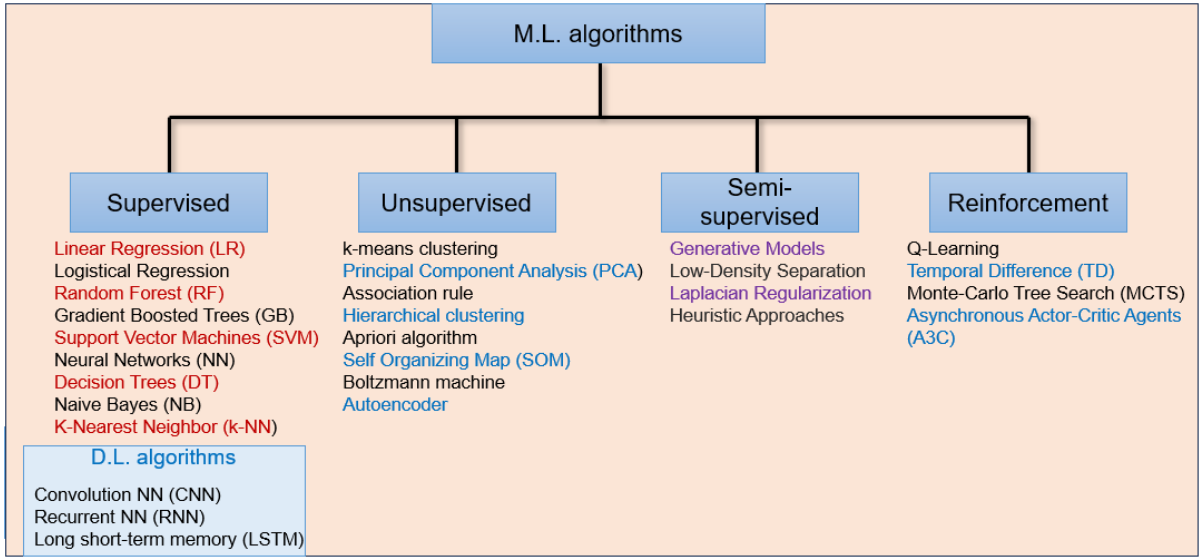

**Figure 7.** Different types of Machine Learning (ML) and Deep Learning (DL) algorithms.

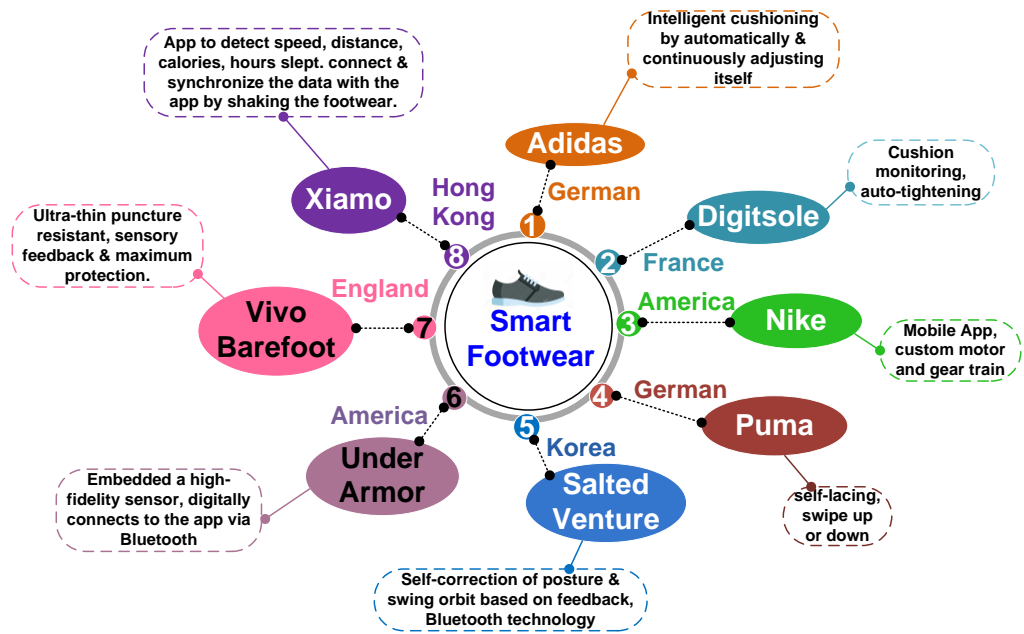

**Figure 8.** Smart footwear companies on the global market.

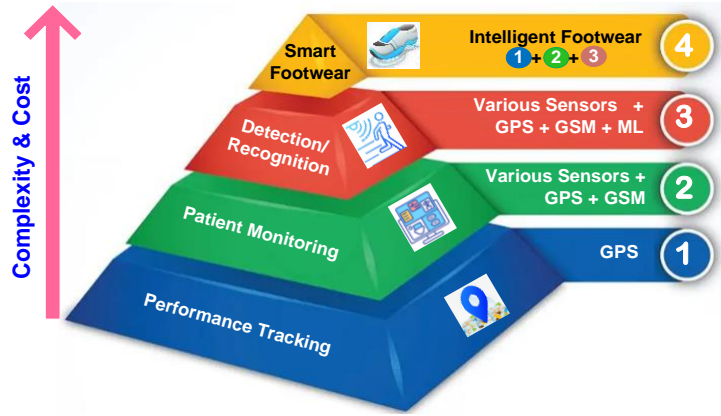

**Figure 9.** Hierarchy for selection of footwear in healthcare

**Table 1.** Suitable sensors for SF design

| Sl. No | Sensor Type, its operation principle, and possible applications in smart footwear |
|---|---|
| 1. | Ultrasonic sensor (3): These sensors utilise ultrasonic waves to measure distance, detect objects and are one of the most commonly used sensors in footwear, specifically for aiding people with visual disability. They can detect the presence or absence of objects within a specific range. Further, these sensors can measure insole thickness and footwear wear & tear and suggest replacement schedules. |
| 2. | LiDAR Sensor (4): Light detection and ranging (LiDAR) based Time of Flight (ToF) sensors are currently the preferred technology for automotive and drone applications. ToF sensors have the emitter, receiver, and processor system on the same PCB/package for easy, cost-effective, and small-footprint integration. They offer high-speed, precise distance measurement independent of target size, colour, and reflectance. The LiDAR sensor can be integrated into the footwear to replace the traditional ultrasonic sensor or added as an additional sensor to support features such as pothole detection, obstacle warning, etc. |
| 3. | Pressure sensor (5): Pressure sensors measure the pressure by converting the applied pressure into an electrical signal that can be measured and utilised for various applications.<br>• Strain Gauge Pressure Sensors: measure the strain or deformation caused by pressure.<br>• Capacitive Pressure Sensors: use changes in capacitance to measure pressure.<br>• Piezoelectric Pressure Sensors: utilise the piezoelectric effect, where pressure generates an electric charge in certain materials.<br>• Resonant Pressure Sensors: use the change in the resonant frequency of a vibrating element under pressure to determine the applied force. They measure pressure at different points on the foot [placed in soles].<br>Additionally, strain gauge sensors installed on the footwear can detect sudden bends and movements in the footwear. Also, pressure sensors can be used to measure weight. |
| 4. | Accelerometer and Gyroscope (6): An accelerometer measures linear acceleration and can detect the movement of an object in terms of acceleration, deceleration, or changes in direction. On the other hand, a gyroscope measures angular velocity around a particular axis. It detects changes in orientation or rotational movements, such as tilting, rotating, or twisting. Inertial measurement units (IMUs) incorporate accelerometers and gyroscopes into a single sensor package, providing a more compact and integrated solution for |
| | motion-sensing applications. IMUs are integrated into wearable devices to monitor and analyse physical activities. They can measure steps, distance, speed, and calories burned and provide feedback on movement patterns and exercise techniques. |
| 5. | Sweat Sensors (7): Skin-worn biosensors can analyse the wearer's sweat to monitor various physiological conditions. The biomarkers in the sweat can be used to detect certain genetic conditions. Also, using the glucose-level correlation between sweat and blood leads to potential applications in the continuous monitoring of diabetes. |
| 6. | Temperature sensor: It detects and measures the hotness and coolness of air, liquids, or solid surfaces and converts them into electrical signals. Types of temperature sensors include: <br><br> a) Resistance Temperature detector (RTD) sensors measure temperature by changing resistance proportional to the temperature. These are available as individual sensors or fully packaged assemblies consisting of a sensor element, a covering, an epoxy or filler, extension leads, and sometimes connectors, thus allowing for flexibility to design (8). <br><br> b) Negative Temperature Coefficient (NTC) thermistors use the properties of ceramic/metal composites that have an inverse relation between resistance and temperature to measure the temperature. NTC sensors have a small size, excellent long-term stability, high accuracy, and precision (9). <br><br> c) A thermocouple sensor is formed by joining two individually insulated dissimilar metals at one end. The temperature is measured at this joint. When the joint is placed in a high-temperature environment, a small voltage is produced at the open ends of the two metals, which can be measured and interpreted (10). <br><br> d) Thermopile infrared (IR) temperature sensors generally provide non-contact temperature measurements. These are composed of small thermocouples on a silicon chip that absorb the incident IR energy and produce an equivalent output signal. Modern sensors also have an integrated reference sensor for calibration and compensation. The output can be an analog voltage or a digital value (11). <br><br> e) Digital Temperature Sensors are IC temperature measurement systems. The miniature packages are designed specifically for tight spaces. The integrated microcircuit design allows quick response to changes in process temperature, fast conversion times, and very low power consumption (12). <br><br> The temperature sensors can be used in smart footwear to keep track of foot temperature and can act as surrogate marker of the activity. Temperature changes in specific areas of the foot can indicate issues such as inflammation, injury, or conditions like diabetic foot ulcers. Elevated temperatures in localized regions might signify inflammation or infection. They can also assist in monitoring blood circulation in the feet. Changes in foot temperature might indicate poor circulation. |
| 7. | Gas sensors can be used to detect foot odour, and they can detect Bromodosis, caused possibly due to fungal infection (13). Bromodosis is smelly feet, and it is often caused by the interaction of sweat with bacteria on the skin's surface. Fungal infections like athlete's foot or other dermatophyte infections can also contribute to foot odor. Gas sensors can detect the specific gases emitted by these fungi, aiding in the early identification of such infections. |

**Table 2.** Technical insights and overview of SF design for patient monitoring

| Ref. No. | Target application | Technical details | Main findings |
|---|---|---|---|
| (47) | Inertial and plantar pressure measurement | Insole, Wrist band, Accelerometer Gyroscope, Pressure sensor, BLE, smartphone, Sampling rate 50Hz. | Best body part for HAR: Feet or wrist. |
| (48) | Six ambulation activities detection | Smart insoles from IEEE Luxembourg SA: Accelerometer, Gyroscope, magnetometer, ECU, BLE, ML algorithms, smartphone, Sampling rate: 200Hz, Recording time:120 minutes, Age group: 25 – 55 years. | Inertial sensors are reliable for dynamic activities; pressure sensors are reliable for stationary activities. |
| (49) | Various motion activities detection | Inertial sensors, DL (CNN). | The use of five inertial sensors was found to be promising. |
| (51) | Foot pressure distribution | Capacitive sensor, ML. | ML can provide the required pressure measurement. |
| (52) | Plantar pressure distribution and daily activities recognition | Seven pressure sensors, FFT, ML algorithms, Sampling rate: 100 Hz, 12-bit resolution, Age group: 26 ± 9 years. | Generalization of the present study is required for large populations and behaviour. |
| (53) | Foot pressure distribution, motion activities | Two hundred eighty capacitive pressure sensors, 56 capacitive temperature sensors, FT, and wired communication. | The smart insole may be an alternative to recognise daily activities. |
| (54) | Plantar pressure distribution – daily activity | MWCNTs/PDMS piezoresistive nanocomposites, LABView. | Used for disease detection and diagnosis purposes. |
| (55) | Seven Types of daily activities recognition | Accelerometer, DL, wireless mode. | SF is one of the easy-to-use devices – for all age groups. |
| (56) | Daily activities recognition | - | 3-D insole might have favourable effects on foot health in a healthy population. |
| (57) | Locomotor activities | Accelerometer, gyroscope, magnetometer (smartphone sensors), FFT, CNN. | User independent system can be built for HAR. |
| (61) | Monitoring pressure, temperature, and humidity of diabetic feet | Two temperature and humidity sensors, eight pressure sensors, BLE, smartphone, alerting system, Arduino 328, Age group: 25-55 years. | Such systems improve patients' self-management and health outcomes. |
| (62) | Device for DFU prevention. | Flexible 2 mm thick insoles with a matrix of 99 capacitance-based sensors, Sampling rate: 50 Hz, resolution: 2 sensors/cm$^2$, a database of 919 diabetic patients. | The pre-clinical studies ensured that the device met user needs and requirements. |
| (63) | Monitoring and preventing DFU | Smartphone plug-and-play, temperature, humidity, and pressure sensors. | DFU can be effectively monitored using a smartphone. |
| (64) | Monitoring DFU: plantar pressure measurement | Pressure sensors, ESP8266, WIFI mode, smartphone. | Appropriate placement of pressure sensors can detect |

| | | | injury and can monitor DFU. |
|---|---|---|---|
| (65) | Monitoring DFU.: plantar pressure measurement | Eight pressure sensors, sensor pod, smartwatch, sampling rate: 8 Hz, age group > 18 years. | Continuous plantar pressure monitoring and dynamic offloading guidance can reduce DFU site recurrence. |
| (66) | Monitoring DFU.: temperature, pressure, humidity measurement | Temperature and humidity sensor, two pressure sensors, microcontroller, WIFI, smartphone. | DFU can be reduced significantly by incorporating sensors inside traditional footwear. |
| (67) | Plantar pressure measurement: monitoring DFU. | Eight capacitive sensors, two commercial flexible PCB and a dielectric sheet, BLE, PC, microcontroller, sampling rate: 100Hz, resolution: 28 bits. | Using such a system improves the efficiency of studying conditions associated with diabetic foot. |
| (69) | Plantar pressure ease: DFU monitoring | Pressure sensor, PC, sampling rate: 50 Hz, 76 participants. | Optimization in terms of length of time is required for real-time use. |
| (70) | Diabetic foot monitoring: temperature measurement | Four temperature sensors, 35 participants. | Continuous monitoring provides preventative information on foot ulcers. |
| (71) | Plantar pressure measurement and foot deformation analysis in case of DFU. | Nineteen female participants, aged 57-75 years, 4D scanner images. | Custom-fabricated insoles and heel pads made of a plurality of materials can help to redistribute the plantar pressure. |
| (73) | Temperature and humidity measurement: DFU. monitoring | Textile fabricated insole: knitted spacer fabric with silicon tubes inlay, leather, five sensors, 21 female participants, age group: 21-30 years. | Thermal comfort can be enhanced with the use of a textile-fabricated insole. |
| (80) | Balance ability evaluation and gait patterns analysis in older women | Thirty older women, 65 – 83 years old, series of laboratory tests, 4 mm thick ethyl vinyl acetate with dome-shaped projections. | Resulted in a significant reduction in step width while walking. |
| (81) | Introduction to smart insole dataset, gait, and PD analysis | Pair of pressure sensors, Accelerometer, 29 participants, sampling rate: 100Hz, 20-59 years age group. | The dataset can be used for the detailed analysis of gait since data have been evaluated by a neurologist specialized in movement disorders. |
| (82) | Fall detection in the elderly (with walking disorder) | Arduino Nano microcontroller, two proximity sensors, a water sensor, a light sensor, a force sensor, a buzzer, and a vibration motor. | smart shoes with intelligent devices need to detect and prevent falls at the same time. |
| (83) | Mobility and gait assessment | Three Force Sensing Resistors, Inertial Measurement Unit (IMU) sensor, Ultrasound sensor on each leg, Arduino controller, BLE. | Abnormalities in walking patterns can be detected. |
| (85) | Flat feet detection | Three force sensors, Accelerometer, BLE, Arduino Nano. | A cost-effective design can be an alternative to existing motion capture systems. |
| (86) | Real-time monitoring of gait | Soft insole: capacitance-based pressure sensor, conductive textile, microcontroller, BLE, sampling rate: 100Hz, 15 participants. | The textile-based smart insole can be an alternative to smart shoes. |
| (87) | Real-time gait monitoring | Inertial sensors, inductive wireless | Results showed a high potential |

| | | charging unit, BLE, visual aid, sampling rate: 100Hz, FFT. | for gait diagnostic and gait rehabilitation assessment. |
|---|---|---|---|
| (89) | Portable real-time gait analysis | multi-cell piezoresistive sensor, an IMU, and a logic unit, sampling rate: 500Hz, 6 minutes recording, ML algorithm, 14 participants. | Learning-based methods improve the stride-to-stride gait parameters. |
| (90) | Measuring spatiotemporal gait parameters and centre of pressure trajectories | Eight piezoresistive sensors, microcontroller, WIFI module, logic unit, IMU, sampling rate: 500Hz, 9 participants, MATLAB software. | The system can be used for out-of-the-lab gait analysis of walking and running. |
| (91) | Estimating foot progression angle | inertial and magnetometer sensing units, Accelerometer, gyroscope, microcontroller, wireless charging, sampling rate: 100 Hz, 14 participants, age group: 22-29 years. | The system could be used for knee osteoarthritis in daily life or clinics without specialized motion capture equipment. |
| (92) | Detecting changes in gait by alcohol intoxication | Twenty participants, wireless mode, ML algorithm. | SF can be used for detecting alcohol-impaired gait. |
| (93) | Locomotion monitoring: real-time kinetic measurement | Pressure sensors, IMUs, WIFI, Smartphone, PC, sampling rate: 100 Hz, 9 participants, MATLAB software. | Acceptable matches were achieved for the measured CoPx and the calculated knee joint torques out of 13 movements. |
| (94) | Plantar pressure measurement | Capacitive sensor: silver and cotton, microchip, USB, laptop, BLE. | Gait phases and different patterns can be detected, and the system is bacterial resistance. |
| (95) | Gait analysis tool: PODOSmart® | IMUs: Sensors, 11 participants, age group: 20-49 years, BLE, sampling rate: 208Hz. | Ease of use without technical education. |
| (96) | Evaluating haptic terrain for older adults and PD patients (TreadPort) | Five bladders, PC, VR terrain, WIFI, microcontroller, CAVE display, camera: 60 frames/sec. | Applicable for gait training for walking impediments caused by PD. |
| (97) | Locomotion monitoring: centre of pressure detection | Five textile capacitive sensors, WIFI, sampling rate: 100 Hz, MATLAB software. | Smart wearable sensors can improve quality of life. |
| (98) | Designing and fabricating biomimetic porous graphene flexible sensor: gait analysis | Graphene nanoplates, SBR foam, silver electrodes, microcontroller, BLE. | The system can monitor older and can help in gait training. |
| (99) | Plantar pressure measurement: gait analysis | Twelve Capacitive sensors: copper and polydimethylsiloxane, PIC microcontroller, BLE, PC. | The design offers correct performance behaviour under footfall. |

**Table 3.** Overview of SF data analysis for disease recognition and classification

| Ref. No. | Study objectives | Techniques | Target group | Outcomes |
|---|---|---|---|---|
| (102) | Stride segmentation of accelerometer data, classification of 3 walking patterns. | TinyML, edge computing. | - | Mean stride duration is around 1.1 with a 95% confidence interval. |
| (103) | Gait segmentation method based on plantar pressure only. | Thresholding: moving average, | Six participants: 19-29 years. | The calculated distribution between stance-phase and swing-phase time is almost |

| | | statistical analysis. | | 60%/40% - aligned with literature studies. |
|---|---|---|---|---|
| (104) | Gait classification using feature analysis. | ML algorithms: RF, k-NN, LR, SVM. | Eighteen participants: 22-31 years. | A combination of accelerometer and gyroscope sensor features with SVM achieved the best performance with 89.36% accuracy. |
| (105) | Gait pattern classification. | ML algorithm: NN. | Eleven participants: 22-33 years. | Built-in Accelerometer and gyro sensor gait-pattern classification system can be used without the constraints of a controlled environment. |
| (106) | Detecting thirteen commonly used human movements. | ML algorithms: PCA, k-NN, ANN, SVM. | Thirty-four participants: average age 22.6 years. | The model proved to be effective, with an accuracy of 86%. |
| (107) | Gait pattern classification. | ML algorithm: NN. | Six participants. | The architecture with three nodes provided effectiveness metrics above 99.6%. |
| (108) | Gait pattern classification. | ML algorithms: k-NN, SVM, ELM, MLP. | - | ELM performed better, with an overall accuracy of 93.54%. |
| (110) | Detecting walking behaviour. | ML algorithms: NN, DL (CNN). | Three participants: 26-27 years. | The best performance was achieved with convolutional layered ANN with an average accuracy of 84%. |
| (111) | Gait type classification. | DL (CNN) | Fourteen participants: 20-30 years. | Experimental results for seven types of gaits showed a high classification rate of more than 90%. |
| (112) | Gait abnormality detection. | DL (CNN) | Twenty-one participants: 24-37 years. | Deployment of CNN-LSTM in Nordic nRF52840 can be revisited with model-pruning and post-training quantization. |
| (113) | Walking pattern analysis. | DL | Video frames. | SF can detect any injury the shoe user is suffering from. |
| (114) | Abnormal gait pattern recognition. | DL (LSTM-CNN) | Twenty-five participants: average age 22 years. | Personalised gait classification approach is accurate and reliable. |
| (115) | Recognition of foot pronation and supination. | DL (NN) | 6 participants. | The system can adequately detect the three footprints types with a global error of less than 0.86. |
| (116) | Foot strike pattern classification. | ML algorithms: LR, conditional inference tree, RF. | Thirty participants: 27-41 years. | The system aided in the research and coaching of running movements and obtained the highest classification accuracy of 94% using RF. |
| (117) | Fall detection. | DL (LSTM-CNN) | - | Applying the data augmentation technique improves DL performance in accuracy and precision. |
| (118) | Fall detection. | Statistical tool and algorithm. | Seventeen participants: 21-55 years. | The method demonstrated satisfactory performances providing a maximum accuracy of 97.1%. |
| (119) | Fall detection. | Advanced fall detection algorithm. | Six participants. | Insole could measure walking speed, the distance covered, and the measurement of balance or weight. |
| (120) | Gait analysis and monitoring. | PCA, event detection algorithm. | Four participants. | The new gate metric (eigen analysis) has great potential to be used as a powerful analytical tool for gait disorder diagnostics. |
| (121) | Gait analysis. | - | - | Addressed hip recovery is built on the clinical standard and allows obtaining quantitative parameters directly from the sensors. |
| (122) | Identification and correction for people with abnormal walking patterns. | ML algorithm: DT. | One thousand two hundred fifty data points - 5 classes with 250 data points. | The machine learning approach has a 91.68% accuracy and shows promise for assisting people with Arthritis. |
| (124) | Gait analysis. | Sum of Manhattan distances (SMD). | Three participants. | Smart socks can be an alternate to smart shoes. |
| (125) | Heart rate estimation. | DL (LSTM-CNN) | Fifteen participants. | Significant levels of heart rate estimation could be made using SF. with a correlation of 0.91. |
| (126) | Heart rate and energy expenditure estimation. | DL (CNN) | Ten participants: 20-24 years. | Estimations can be accurate by effectively selecting the optimal sensors. |
| (128) | Gait analysis. | Multivariate analysis, statistical tool. | Twenty-nine participants: 43-75 years. | SF is ideally suited for preoperative evaluation in the clinical setting. |
| (129) | Human behaviour classification. | ML algorithm: RF. | Six participants: 20-22 years. | Four types of behaviours were classified with an F-measure of 78.6%. |

| (130) | Knee abduction movement detection. | ML algorithm: MLP regressor. | One participant: 24 years. | The system performed well in predicting KAM with an accuracy of 87%. However, more experimentation is required. |
|---|---|---|---|---|

**Table 4.** Details of Smart footwear available in the market (137)

| Smart Footwear (Name of the Company) | Applications | Type of the Sensor (No. of sensors used) | Pressure (kPa) | IE | MDAR (Hz) | DTT | BA (Hrs) | Cost |
|---|---|---|---|---|---|---|---|---|
| WIISEL | Continuous gait monitoring, analysis and Fall risk assessment | Piezoresistive (14) | 350 | Yes | 33.3 | BLE | 16 | ---- |
| Pedar (Novel) | Footwear design and injury prevention | Pressure (99) | 600 | No | 100 | BT | 1 | 15,540 € |
| F-Scan (Tekscan) | gait analysis & biomechanics, diabetic offloading, sports medicine | Pressure (960) | 862 | No | 165 | USB Wi-fi | 0.2 | 16,000 $ |
| BioFoot (IBV) | Sports Gait analysis Footwear design | Pressure (64) | 1200 | No | 500 | Wi-Fi | 1 | 12,995€ |
| paroLogg/paroTecc (paromed) | Foot pressure analysis | Pressure (32) Inertial | 625 | No | 300 | Wi-fi | 1.5 | ---- |
| Foot Pressure MS (Medilogic) | Gait, Sports, Health Prevention, Prosthesis and Orthotics, Diabetics | Solid State Relay (SSR) sensors (240) | 640 | No | 300 | Wireless | ---- | ---- |
| Smart Step | Rehabilitation process | ---- | ---- | No | ---- | Card | ---- | 6,000$ |
| Smart Insoles (24eight, LLC) | Medical, sports, and gaming | Pressure (4) Inertial | 241 | Yes | | Wire-less | ~100 | ---- |
| OpenGo science (Moticon) | Medical and sports science Rehabilitation | Pressure (13) Inertial | 400 | Yes | 100 | Wire-less | ---- | 2,000 $ |
| | and training analysis | | | | | | | |
| Footswitches Insole (B&L Engineering) | Gait analysis | Pressure sensors (04) | ---- | No | ---- | Wire-less | ---- | 9,000 $ |
| IE-Integrated Electronics, MDAR-Maximum Data Acquisition Rate, DTT-Data Transfer Type, BA-Battery Autonomy | | | | | | | | |